\documentclass[sn-aps]{sn-jnl}
\usepackage[table]{xcolor}

\usepackage{graphicx}%
\usepackage{multirow}%
\usepackage{amsmath,amssymb,amsfonts}%
\usepackage{amsthm}%
\usepackage{mathrsfs}%
\usepackage[title]{appendix}%
\usepackage{xcolor}%
\usepackage{textcomp}%
\usepackage{manyfoot}%
\usepackage{booktabs}%
\usepackage{algorithm}%
\usepackage{algorithmicx}%
\usepackage{algpseudocode}%
\usepackage{listings}%
\usepackage{comment}
\usepackage{makecell}
\usepackage{ragged2e}

\theoremstyle{thmstyleone}%
\theoremstyle{thmstyletwo}%

\theoremstyle{thmstylethree}%

\begin{document}

\title[Article Title]{\mbox{Beam Manipulation for Terahertz Communications}: A New Quality Productive Force
}

\author[1]{\fnm{Mingxiang} \sur{Li}}\email{mingxiang.li@sjtu.edu.cn}

\author[2]{\fnm{Josep M.} \sur{Jornet}}\email{j.jornet@northeastern.edu}

\author[3]{\fnm{Daniel M.} \sur{Mittleman}}\email{daniel{\_}mittleman@brown.edu}

\author*[1]{\fnm{Chong} \sur{Han}}\email{chong.han@sjtu.edu.cn}

\affil*[1]{\orgdiv{Terahertz Wireless Communications (TWC) Laboratory}, \orgname{Shanghai Jiao Tong University}, \orgaddress{\city{Shanghai}, \postcode{200240}, \country{China}}}

\affil[2]{\orgdiv{Department of Electrical and Computer Engineering (ECE)}, \orgname{Northeastern University}, \orgaddress{ \city{Boston}, \state{MA}, \postcode{02115}, \country{USA}}}

\affil[3]{\orgdiv{School of Engineering}, \orgname{Brown University}, \orgaddress{ \city{Providence}, \state{RI}, \postcode{02912}, \country{USA}}}

\abstract{The terahertz frequency band, ranging from 0.1 to 10 THz, offers extensive spectral resources for next-generation wireless communication systems. To compensate for the limited transmission power of terahertz transceivers and the significant propagation losses in terahertz channels, high-gain directional antennas are essential. Dynamic beam manipulation is therefore crucial for enabling practical communication applications. Moreover, the stringent gain requirements in terahertz systems result in an expanded Fresnel region, highlighting the critical need for efficient beam manipulation techniques in both near-field and far-field conditions. This article provides a comprehensive overview of terahertz beam manipulation techniques. It begins with an introduction of diffraction theory as the foundational propagation model for beam manipulation. Detailed examples tailored to specific communication scenarios are then presented. Experimental verifications using 3-D printed lenses are included for three distinct beam manipulation cases. Alternative approaches for achieving beam manipulation, such as metasurfaces and reconfigurable intelligent surfaces, are briefly discussed.}

\maketitle

\section{Introduction}\label{sec1}

Wireless communication systems have become part of our daily life and are indispensable across industries. Future networks are anticipated to support a wide range of demanding applications, from virtual, augmented, and mixed reality to remote control of sensitive operations \cite{Giordani2020,Chen2021,Akyildiz2022,Jornet2024}. The limited spectral resources in the microwave frequencies and the escalating demand for high data capacity, therefore, have prompted the investigation of unexplored frequency resources in higher frequency bands. The terahertz spectrum offers ultra-wide bandwidth and holds immense potential for future 6G and beyond networks. Frequency range covering 0.1--1 THz can theoretically support transmission speeds up to terabits per second (Tb/s) level \cite{Petrov2020,Nagatsuma2016}. Recent advancements further demonstrate the feasibility of terahertz communications in outdoor scenarios with long-distance wireless transmission \cite{Liu2024,Nagatsuma2024}. These achievements mark a critical step forward for 6G communications, transitioning from laboratory experiments to real-world deployment. \par

However, despite these advantages, terahertz communications face significant challenges, in part due to the lack of high-power solid-state sources \cite{Sirtori2002,Makhlouf2023}, substantial power loss during wireless propagation \cite{ChenYi2024,Han2023}, and limited availability of energy-efficient transceivers \cite{Withayachumnankul2018,Pant2023}.
As the operating frequency increases, terahertz sources suffer from a substantial power decrease as frequencies approaching the terahertz range \cite{Suzuki2024}. The performance of electronic components such as multipliers and amplifiers degrades significantly at terahertz frequencies, mainly due to fabrication constraints and the parasitic effects \cite{Biswas2018}. Conventional CMOS-based devices therefore generally exhibit suboptimal performance in terms of DC-RF conversion efficiency \cite{Ahmad2016,Gao2022}. Heterogeneously integrated devices can achieve higher power, but their increased fabrication complexity and costs limit mass production viability \cite{Mei2015,Arabhavi2022}. For sources above 300 GHz, whether multiplier-chain-based or oscillation-based, the output power of single-source elements typically falls below 1~mW \cite{Jain2020,Gao2022a,Makhlouf2023}. Recent advances, such as a $6\times 6$ resonant-tunneling diode (RTD) array by Canon, Inc., have demonstrated a higher output power over 10~mW at 450 GHz \cite{Koyama2022}. However, without power amplification, the output power will rapidly decrease to $\mu$W levels at higher frequencies \cite{Izumi2017,Aghasi2017}. Photonic generation methods offer an alternative by leveraging optical technologies, typically by photomixing two continuous-wave laser beams at different optical frequencies and down-converting them to terahertz frequencies \cite{Ishibashi2022,Makhlouf2021}. While photonic methods can support higher data rates \cite{Weijie2024,Maekawa2024}, the output power from photodiodes typically remains lower than that of electronic sources \cite{Nellen2020,Grzeslo2023}. \par

The limited power of terahertz signals is subject to strong free-space path loss, which increases with the square of both frequency and distance, as described by Friis' law \cite{Friis1946}. While terahertz transmission also experiences atmospheric attenuation due to interactions with water vapor molecules, multiple low-absorption windows in the terahertz spectrum, such as 220--330 GHz, 625--725 GHz, and 780--910 GHz, have been identified to minimize atmospheric absorption \cite{Nagatsuma2016}. Moreover, recent progress also theoretically prove that such molecular absorption can enhance secure wireless links \cite{Han2023}. \par

Given the limited output power of terahertz sources and high propagation losses, terahertz transmissions are typically highly directional rather than isotropic to counteract these significant losses. Traditional source-integrated metallic antennas, despite their broader half-power beamwidth \cite{Mai2020,Okada2015,Diebold2016,Li2023a,Li2023}, exhibit insufficient gain at terahertz frequencies due to their limited aperture size. Additionally, metal surface roughness induces scattering losses, further degrading the antenna performance \cite{Luk2017}. To enhance the directivity, additional antenna designs such as lens antennas and reflector dishes are often incorporated to expand the aperture \cite{Llombart2011,Li2023a,HHI2024}. Traditional lens antennas and parabolic mirrors can be fabricated through CNC machining or 3-D printing. Although materials like Teflon or other polymers are cost-effective and provide good impedance matching with the sources, their high dielectric losses pose limitations \cite{Yi2016,Konstantinidis2017,Wu2019a}. High-resistivity silicon paired with matching layers offers a promising alternative for terahertz antennas, but challenges in fabrication techniques, such as silicon micromachining and deep reactive ion etching (DRIE), make it difficult to achieve both compactness and high aperture efficiency \cite{Liang2021,Dechwechprasit2023,Lees2024,Campo2020,Alonso-delPino2024,Li2024a}. This underscores the need for advanced development of compact, efficient terahertz transceivers. \par

\begin{figure}[!t]
\centering
\includegraphics{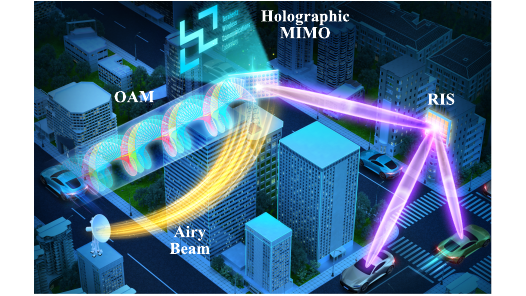}
\caption{Beam manipulation for terahertz communications in a comprehensive environment.}
\label{fig_Fig1}
\end{figure} 

Considering the aforementioned motivations and constraints, terahertz communications rely on establishing highly directional, narrow-beamwidth beam linkages between devices. Such setup demands dynamic beam manipulation to enable flexible switching between devices. Achieving the high gain required in terahertz transceivers necessitates a larger aperture dimension $D$, which extends the radiative near-field region. This region is typically known as the Fraunhofer distance, defined as $2D^2/\lambda$, where $\lambda$ is the wavelength. Within this region, wavefronts can exhibit arbitrary shapes, in contrast to the planar wavefronts characteristic of the far-field. Unlike traditional wireless systems that mainly operate in the far-field, practical terahertz networks will likely encompass both near- and far-field conditions, necessitating comprehensive beam manipulation strategies in the novel paradigm of cross near- and far-field regions \cite{Liu2023,Wang2024}. Promisingly coming into the near-field, doors are opened to many interesting types of beam manipulation, as shown in Fig. \ref{fig_Fig1}. Recent growing interest in beam engineering for millimeter- and terahertz-wave frequencies, particularly related to ultra-massive multiple-input multiple-output (UM-MIMO) systems \cite{Chen2023,Han2024}, also highlights the importance of developing adaptable beam manipulation techniques to support both near- and far-field communications in the terahertz band.

\section{Propagation model for beam manipulation}\label{sec2}

To understand the mechanism of beam manipulation, one must first understand how electromagnetic (EM) wave propagates. Maxwell's equations provide a systematic model for general EM wave characteristics, showing how EM waves propagate through complex environments, including reflections, diffractions, and other interferences. To simplify this, a common practice is to consider EM waves as vector fields with amplitude and phase, propagating in a homogeneous medium such as air, comparable to light travelling in free space in optics. Here, for simplicity we temporarily neglect polarization effects. Before Maxwell's equations, physicists in optics proposed speculative propagation models. The Huygens-Fresnel principle builds on these early ideas by proposing that every point on a wavefront can be treated as a secondary source with spherical wavelets. These wavelets interfere with one another, and their cumulative effect forms a new wavefront. As a scalar theory, the Huygens-Fresnel principle rigorously calculates the intensity and phase of an electric field in scalar diffraction theory, thus modelling the propagation behavior of EM waves effectively \cite{Huygens}. \par

Various mathematical models and their approximations and derivations based on the Huygens-Fresnel principle can be found in beamforming or wavefront engineering articles and tutorials \cite{Huygens,Headland2018,Singh2024}. Here, we provide only a summary: assuming a complex electric field at the initial aperture $E(x_0,y_0,0)$, the corresponding complex electric field at a propagation distance {\it d} can be described by the following expression: 
\begin{equation}
E(x_1,y_1,d) = \frac{-j}{\lambda}\sum_{x_0}\sum_{y_0} E(x_0,y_0,0) \frac{\exp(-j{k}r_{01})}{r_{01}} \frac{1+(d/r_{01})}{2} \Delta x_0 \Delta y_0,
\label{Eq:HuygensPrinciple}
\end{equation}
where $\lambda$ is the wavelength, $k$ is the wavenumber, and $r_{01}$ is the distance between source point ($x_0,y_0,0$) and the observation point ($x_1,y_1,d$), expressed as: $r_{01} = \sqrt{(x_1 - x_0)^2 + (y_1 - y_0)^2 + d^2}$. It is important to note that this expression involves a computationally intensive process, as each point on the targeted transverse plane requires summation over a large array of secondary sources. To simplify the analysis, approximations for the near-field and far-field can be employed using Fresnel and Fraunhofer diffraction equations, respectively. By implementing fast Fourier transform (FFT), these approximations offer more computationally efficient approaches.\par
\begin{figure}[!t]
\centering
\includegraphics{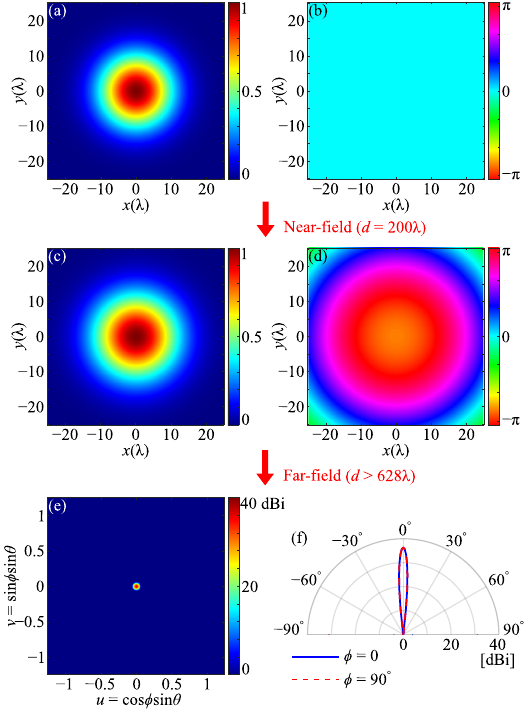}
\caption{Beam propagation model for a collimated Gaussian beam. (a, b) Initial amplitude and phase distributions at the source plane. (c, d) Amplitude and phase distributions after a propagation distance of 200$\lambda$ (approximately one-third of the Fraunhofer distance) in the near-field. (e) Orthographically projected 2-D radiation pattern. The maximum range of {\it u} and {\it v} is determined by the resolution, and the range where $u^2+v^2 > 1$ does not have any physical meaning, as they fall outside the beam's angular extent. (f) 1-D radiation patterns. }
\label{fig_Gau}
\end{figure} 
Coming into the near-field, where the observation plane is close to the source aperture, Fresnel diffraction provides an expression for the complex electric field in Cartesian coordinates. This can be reformulated in a convolution form as: 

\begin{equation}
E(x_1,y_1,d) = E(x_0,y_0,0)*h(x_0,y_0,d), 
\label{Eq:Fresnel}
\end{equation}
where $h(x_0,y_0,d)$ is the impulse response of light propagation and can be expressed as:
\begin{equation}
h(x_0,y_0,d) = {\exp(jkd}) \cdot \frac{1}{j\lambda d} \cdot {\exp(j\pi\frac{{x_0}^2+{y_0}^2}{\lambda d}}).
\label{Eq:Transfer}
\end{equation} 

In the far-field, where the observation distance is sufficiently large, Fraunhofer diffraction simplifies the analysis. Here, a two-dimensional Fourier transform is applied to the complex electric field:
\begin{equation}
E(u,v) = \mathcal{F} \{E(x,y,0)\}.
\label{Eq:Fraunhofer}
\end{equation}
In this form, the 3-D far-field radiation pattern is represented as an orthographically projected 2-D radiation pattern in terms of {\it u} and {\it v}, which are expressed as $u = \mathrm{cos}(\phi)\mathrm{sin}(\theta)$ and $v = \mathrm{sin}(\phi)\mathrm{sin}(\theta)$, respectively. Here, $\phi$ and $\theta$ are the azimuth and elevation angles. Note that {\it u} and {\it v} can be derived from the spatial-frequency components as $u = {k_x}/{k}$ and $v = {k_y}/{k}$. \par

Using the aforementioned propagation models, we can analyze the behavior of a typical beam with Gaussian amplitude (i.e., beam radius $w_0 = 10\lambda$) and uniform phase distribution, as depicted in Figs. \ref{fig_Gau}(a, b). The analysis explores the beam's propagation over a distance of $200\lambda$ in the near-field through Eqs. (\ref{Eq:Fresnel}, \ref{Eq:Transfer}) and examines the resulting far-field patterns through Eq. (\ref{Eq:Fraunhofer}). As shown in Figs. \ref{fig_Gau}(c, d), as the beam travels in free space, the beam waist expands in the near-field and the initial planar wavefront gradually transforms into a spherical shape. Since the beam starts with a planar phase profile, it exhibits a collimated wavefront, resulting in a high-gain performance. This property leads to a narrow beamwidth in the far-field, as shown in Fig. \ref{fig_Gau}(e). The corresponding 1-D far-field patterns at two planes where angles $\phi = 0^\circ$ and $\phi = 90^\circ$ are provided in Fig. \ref{fig_Gau}(f).

\section{Concept of beam manipulation}\label{sec3}

In section \ref{sec2}, we establish the propagation model of a given beam, which can be determined by its original complex {\it E}-field distribution over a transverse plane. An example of a Gaussian beam with a uniform phase profile propagating in free-space was provided. In this section, specific desired propagating beam behaviors can be achieved through manipulation of the complex {\it E}-field distribution. While in conventional microwave or optics systems, both amplitude and phase manipulation are commonly employed, the situation is different in the terahertz range. Due to the limited power budget in terahertz systems, controlling the amplitude results in significant power losses \cite{Ayyagari2023,RIS_Wang2024}, placing a heavy burden on the power budget. Moreover, for a terahertz MIMO systems, phase shifting can be achieved by each element but amplitude control can be achieved only with a large number of elements as a group \cite{Akyildiz2016,Ning2023}. As a result, amplitude control is less common in terahertz applications. Instead, phase manipulation is the more widely used approach.

Section \ref{sec3.1} primarily focuses on phase manipulation, assuming a fixed Gaussian amplitude with a beam radius of $w_0 = 10\lambda$ for the source. Although less common, section \ref{sec3.2}  briefly discusses beam control involving both amplitude and phase, and section \ref{sec3.3} provides a comparison. To simplify the analysis, specific operating frequencies are not provided, and the axis will be represented in terms of wavelengths. For clarification, in literature the {\it y}-axis is sometimes defined as the beam propagation direction, since the {\it z}-axis is pointing perpendicular to the ground. In this article, we define the transverse plane by {\it x}- and {\it y}-axes, and the {\it z}-axis as the beam propagation direction. The resolution of the Cartesian coordinates will be set at 0.2$\lambda$ to ensure a sufficiently small secondary point source. Additionally, practical implementations in communication scenarios will be introduced to illustrate how beam manipulation can aid in terahertz communications.

\subsection{Phase manipulation}\label{sec3.1}

\subsubsection{Focused beam}\label{sec3.1.1}

\begin{figure}[!t]
\centering
\includegraphics[scale=0.7]{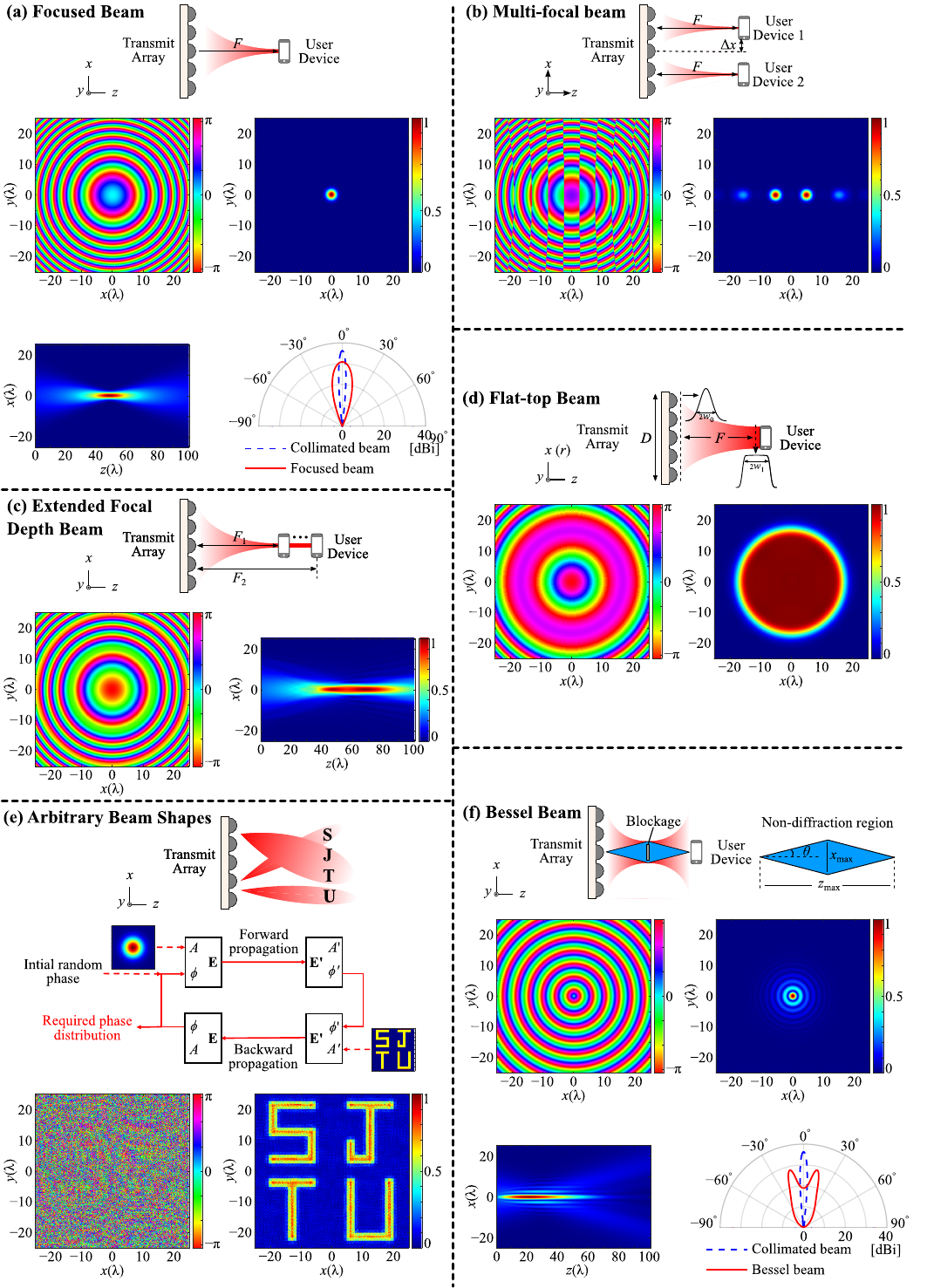}
\caption{Beam manipulations. (a) Focused beam. (b) Multi-focused beam. (c) Extended-focal-depth beam. (d) Flat-top beam. (e) Beam shape with letters ``SJTU". (f) Bessel Beam. }
\label{fig_full_1}
\end{figure}

In the first case, we consider a device that requires maximum power reception, placed at the center of the transverse plane and within the near-field from the transmitter, as shown in Fig. \ref{fig_full_1}(a). To maximize the energy at a specific spot in the near-field, a focused beam is necessary. To focus the beam at a point at a distance {\it F} from the radiating aperture, the beam's phase distribution can be employed as follows:  
\begin{equation}
\phi(x,y) = k_{0}(\sqrt{F^{2}+(x^{2}+y^{2})} - F).
\label{Eq:Focus}
\end{equation}
An example of this phase distribution, corresponding to a focal length of $50\lambda$, is shown in Fig. \ref{fig_full_1}(a). The resulting beam propagation profile at the focal distance and cross-sectional view are depicted are also given. As demonstrated, the radiation from the aperture converges to a small focal point. In addition to communications applications, such focused beams are commonly used in imaging and sensing systems \cite{Lee2020,Dobroiu2022,Chung2024}, where the small focal spot provides exceptionally high resolution. The smaller diffraction limits enable the detection and resolution of finer details. On the other hand, the focusing performance in the near-field will essentially lead to quicker beam divergence, resulting in a reduced gain performance and an increased beamwidth in the far-field, as shown in the radiation pattern.

\subsubsection{Multi-focal beam}\label{sec3.1.2}

In a following scenario, we consider two devices requiring maximum power reception, positioned symmetrically at a distance from the center and still at distance {\it F} from the radiating aperture, as depicted in Fig. \ref{fig_full_1}(b). To focus the beam at one focal point offset by $\pm \Delta x$ from the center, the phase distribution can be adjusted as follows:
\begin{equation}
\phi(x,y) = k_{0}(\sqrt{F^{2}+((x\mp \Delta x)^{2}+y^{2})} - F).
\label{Eq:Focus}
\end{equation}
The superposition of two complex electric fields corresponding to separate focal points can help achieve multi-focal behavior. Figure \ref{fig_full_1}(b) also shows the resulting phase profile and demonstrates that the beam is shaped to target the desired locations. Notably, the superposition modifies the amplitude distribution, but in practice, the source amplitude often retains its original Gaussian distribution. This mismatch introduces imperfections, such as sidelobes, as can be observed in Fig. \ref{fig_full_1}(b). The method of simple superposition is an efficient approach for generating beams with a limited number of focal points. For scenarios requiring beams with numerous focal points, more complex methods, such as involving machine learning assisted algorithm, are recommended \cite{Komorowski2021,Komorowski2024}.

\subsubsection{Extended-focal-depth beam}\label{sec3.1.3}

In a similar scenario, we consider a single device requiring maximum power reception, positioned at the center but possibly within a range from $F_1$ to $F_2$, as depicted in Fig. \ref{fig_full_1}(c). To ensure that the device receives maximum power across this range along the {\it z}-axis, the beam's focal depth must be extended. Similarly, this extended focal depth can be achieved by superimposing the complex electric fields corresponding to focal points at $F_1$ and $F_2$, and the phase distribution can be adjusted accordingly. An example of the phase distribution required for a focal depth from around $50\lambda$--$70\lambda$ is given in Fig. \ref{fig_full_1}(c). \par

As illustrated in Fig. \ref{fig_full_1}(c), the focal depth along the {\it z}-axis is effectively extended, allowing for consistent power delivery over the specified range. This extended focal depth behavior is also highly advantageous for imaging applications. For instance, conventional raster scanning methods typically provide a 2-D image at a precisely-fixed distance and require physical movement of the object for depth exploration \cite{Chung2024,Hu2024}. By using a beam with an extended focal depth, the object can be placed within a loose range and potentially, a 3-D image of an object can be captured without needing to reposition the object \cite{Jiang2013,Hernandez-Serrano2021,Jia2022}.

\subsubsection{Flat-top beam}\label{sec3.1.4}

In the next case, we consider a single device with a large receiving aperture, as depicted in Fig. \ref{fig_full_1}(d). To maximize power coupling efficiency \cite{Hoogelander2024,Li2024b}, it is ideal to ensure uniform power distribution across the receiver aperture. In other words, a flat-top beam is preferred over the conventional Gaussian beam. To transform a Gaussian beam with a beam radius of $w_0$  into a flat-top beam with a beam radius of $w_1$ at a propagation distance $F$, a common approach is to employ a phase distribution determined through the geometrical transformation technique \cite{Abbaszadeh2019,Abbaszadeh2020}.Since the profile is radially symmetric, we provide the 1-D expression in a radial form:
\begin{equation}
\phi(r) = \frac{2\pi}{\lambda F} \int_{-D/2}^{r} \left[ 2w_1 \left( \frac{f(r')}{f(D/2)} - \frac{1}{2} \right) - r' \right] dr',
\label{Eq:Flat_top1}
\end{equation} 
\begin{equation}
f(x) = \int_{-D/2}^{x} \exp\left(-\frac{2\zeta^2}{w_0^2}\right) d\zeta,
\label{Eq:Flat_top2}
\end{equation}
where {\it D} is the dimension of the radiating aperture. Here, we provide an example of a conversion to a flat-top beam radius $w_1 = 15\lambda$ at a propagation distance of $50\lambda$. The resultant phase profile is optimized and illustrated in Fig. \ref{fig_full_1}(d). Alternative approaches, such as iterative techniques like the Gerchberg-Saxton algorithm, are also available for determining the necessary phase distribution \cite{Li2023b}. This phase profile modifies the Gaussian beam's intensity distribution, transforming it into a flat-top beam with uniform amplitude over the specified aperture. Such uniform illumination finds extensive applications in super-resolution imaging \cite{Stehr2019}, and holography systems \cite{Kleindienst2010}. It is important to note that in the microwave community, a flat-top beam typically refers to a far-field concept characterized by uniform gain performance over a certain angular range \cite{Nguyen2011,Singh2019}. However, in this article, we focus on achieving uniform illumination on a planar screen. Notice that a flat-top beam in the near-field does not necessarily maintain a flat-top profile in the far-field.

\subsubsection{Arbitrary beam shapes}\label{sec3.1.5}

Beyond flat-top beams, Gaussian beams can be transformed into various arbitrary beam shapes, a technique commonly referred to as computational holography \cite{Jia2022} or holographic MIMO \cite{Huang2020,Gong2024}, as illustrated in Fig. \ref{fig_full_1}(e). To generate arbitrary beam shapes, phase retrieval algorithms are typically employed. One widely known approach is the iterative algorithm introduced by Gerchberg and Saxton in 1972, commonly known as the GS algorithm \cite{Gerchberg1972}, which retrieves the phase of a source plane to achieve a desired target intensity over another plane. The GS algorithm follows these steps: \par

\begin{itemize}
    
\item{Initialization}: Start with a random phase distribution combined with the source amplitude distribution.\par
\item{Forward propagation}: Propagate the wavefront forward to generate a complex wavefront in the target plane.\par
\item{Target amplitude replacement}: Replace the resultant amplitude in the target plane with the desired target amplitude distribution while retaining the phase.\par
\item{Backward propagation}: Propagate the new wavefront back to the source plane.\par
\item{Source amplitude replacement}: Replace the amplitude at the source plane with the initial source amplitude while retaining the updated phase. \par

\end{itemize}

These steps are iterated until convergence is achieved, at which point the algorithm produces the required phase distribution for generating the target beam shape. A detailed schematic of this process is shown in Fig. \ref{fig_full_1}(e). One significant limitation of the GS algorithm is its slow convergence after some initial iterations. Due to starting with a random phase distribution, this might lead to convergence to suboptimal phase distributions, potentially resulting in inaccuracies. To address this, various improvements have been proposed, such as the Yang-Gu algorithm \cite{Yang1994} and the multiple-stage phase retrieval algorithms \cite{Rodrigo2010,Zhao2010}. These enhancements introduce constraints during the transformation process to improve the convergence of the algorithm and the image quality. It should be clarified that conventional GS algorithm is applied when the target plane is in the far-field, therefore the propagation model is a simple Fourier transform. In a near-field target plane case, as discussed in Section \ref{sec2}, the Fourier transform is replaced with the Fresnel transform to account for near-field propagation effects. Recently, machine learning-assisted phase retrieval algorithms have gained attention for their ability to improve efficiency and accuracy \cite{Rivenson2019,Huang2023}. These methods leverage advanced optimization techniques to overcome the limitations of traditional iterative approaches, offering promising results in beam shaping and related applications \cite{Liao2022,Xiang2024}. \par

Here, an example of transforming a Gaussian beam into a beam shape containing the letters ``SJTU" (Shanghai Jiao Tong University) after a propagation distance of $50\lambda$, is shown below. The phase profile is derived by a modified GS algorithm \cite{Li2023b}, and the resulting field distribution is displayed in Fig. \ref{fig_full_1}(e). It should be noted that the algorithm used here is for demonstration purpose and does not represent an optimal implementation. Since the transformation happens in the near-field and certain beam shapes only last for a very short propagation distance, no far-field results will be provided in this case. Nevertheless, such beam-shaping techniques are also applicable in the far-field to generate customized radiation patterns \cite{Cai2024}.

\subsubsection{Bessel beam}\label{sec3.1.6}

In previous cases, we mainly considered line-of-sight communication scenarios. However, in practical situations, obstacles that block the line-of-sight are common, as illustrated in Fig. \ref{fig_full_1}(f). When the obstacle size is relatively small, one potential solution is to utilize a Bessel beam. The amplitude distribution of a Bessel beam can be described by a Bessel function of the first kind, therefore the cross-section of a Bessel beam consists of concentric rings. While an ideal Bessel beam is theoretically infinite and non-diffractive, it is impractical to generate such a beam due to the requirement of infinite energy. Instead, in real-world applications, Bessel beams are non-diffractive only within a finite propagation range. Bessel beams also exhibit a unique self-healing property. Because the beam comprises concentric rings, if a portion of the beam is obstructed by a small object, the pattern will regenerate beyond the obstruction \cite{Katsuue2023,Reddy2023}. To generate a Bessel beam with a deflection angle $\theta$, the phase distribution required can be expressed as:
\begin{equation}
\phi(x,y) = k_{0}\sqrt{x^{2}+y^{2}}\mathrm{sin}\theta.
\label{Eq:Bessel}
\end{equation}
An example of this phase distribution, corresponding to a deflection angle of $13.8^\circ$, is illustrated in Fig. \ref{fig_full_1}(f). The resulting beam profile at a propagation distance of $30\lambda$ and the cross-sectional view are shown in Fig. \ref{fig_full_1}(f). As depicted, a prominent column of constructive interference can be observed along the {\it z}-axis, which serves as the primary functional component of the beam. Surrounding this central region are fringing fields that form concentric rings. These rings are an interference pattern created by the superposition of waves. The maximum non-diffraction region is primarily determined by the original aperture size (e.g., $w_0$ for the Gaussian beam) and the deflection angle $\theta$. The maximum distance for the non-diffraction region along the {\it z}-axis can be approximated as: $z_{\mathrm{max}} = w_0/\mathrm{tan}(\theta)$, which is around $42\lambda$. The non-diffraction region width approximately equals to $w_0$, which is around $10\lambda$ along the {\it x}- and {\it y}-axes. Since the deflection angles are applied symmetrically across the aperture, the far-field pattern shows that the main lobes are located at the deflection angle $\theta$. The gain at the broadside direction is therefore reduced from the collimated Gaussian beam case.

\subsubsection{Airy beam}\label{sec3.1.7}

\begin{figure}[!t]
\centering
\includegraphics[scale=0.7]{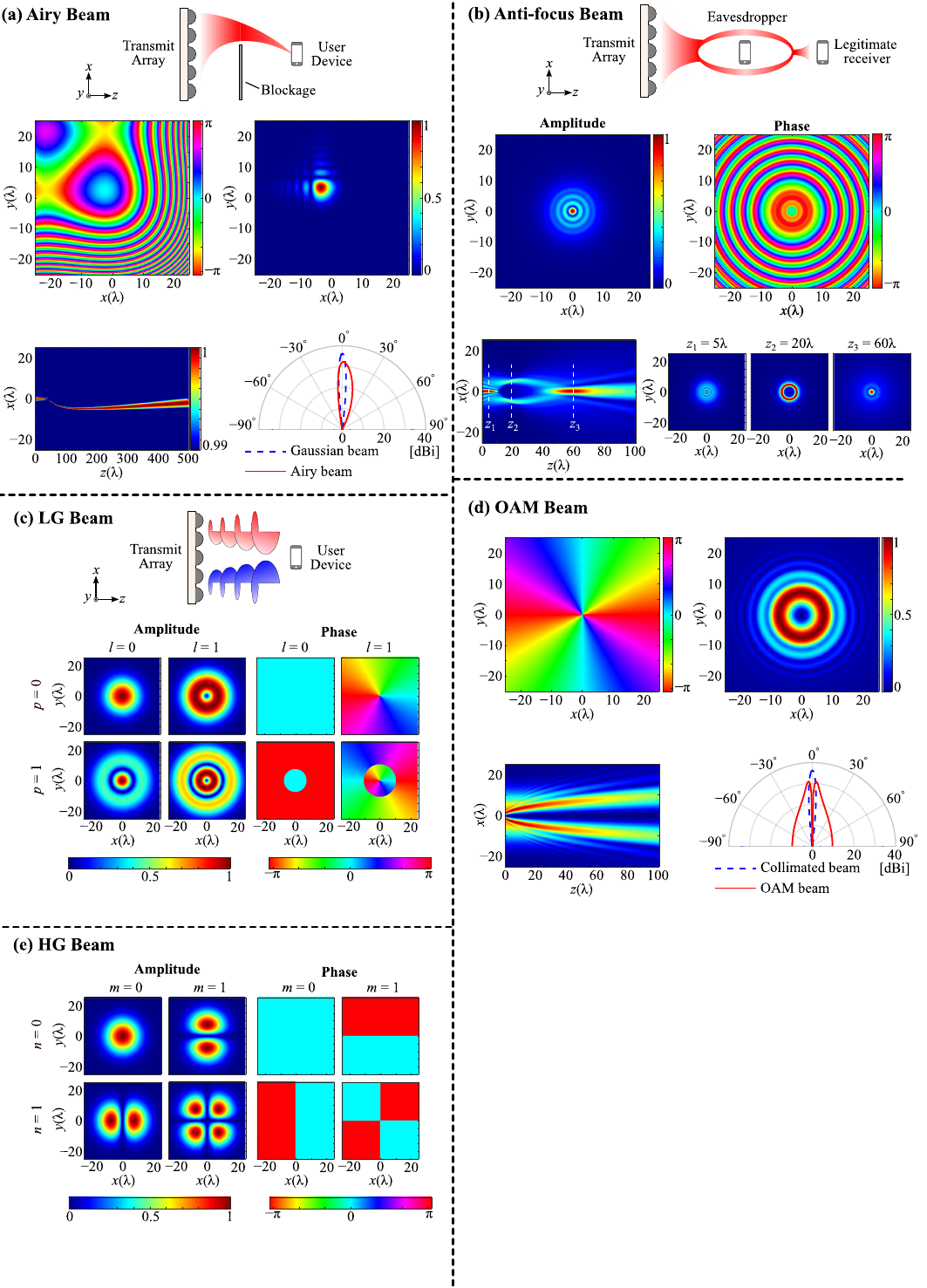}
\caption{Beam manipulations. (a) Airy beam. (b) Anti-focused beam. (c) LG beams (d) OAM beam. (e) HG beams.}
\label{fig_full_2}
\end{figure} 

In the next case, we consider a scenario with a larger blockage, as illustrated in Fig. \ref{fig_full_2}(a). As discussed in the section on Bessel beams, the non-diffraction region is limited to a specific range. When the obstacle size becomes relatively large, Bessel beams lose their ability to reform their shape. A potential solution in such situations is to utilize an Airy beam with its unique property of self-acceleration. As the Airy beam propagates, it bends and follows a curved parabolic trajectory. The central intensity, or the main lobe, adheres to this parabolic path, accompanied by several smaller side lobes located along one side of the parabola \cite{Efremidis2019}. Theoretically, Airy beam can be generated with an amplitude distribution following an Airy function \cite{Guerboukha2024,Petrov2024}, but hard for practical implementation. To transform a Gaussian beam into an Airy beam, the most common method is to apply a cubic phase distribution \cite{Miao2020}, expressed as:
\begin{equation}
\phi_\mathrm{airy}(x,y) = a ({x^{3}+y^{3}}),
\label{Eq:Airy}
\end{equation}
where $a$ serves as a scaling factor that determines the bending of the trajectory, effectively governing the size of the obstacle the beam can bypass. It is important to note that a larger scaling factor results in a larger curvature but less confined side lobes, which reduces the gain performance in the far-field. To twist the beam back towards the center in the near-field, an additional quadratic phase distribution $\phi_\mathrm{focus}(x,y) = b ({x^{2}+y^{2}}$), where $b$ is another scaling factor that determines focal length, can be added to the phase profile. This adjustment effectively rotates the trajectory of the beam to ensure it aligns with the receiver's position in the targeted position. The final phase distribution can be expressed as the summation of $\phi_\mathrm{airy}(x,y)$ and $\phi_\mathrm{focus}(x,y)$. \par

An example of the final phase distribution, combining linear steering, quadratic focusing, and cubic curving components, is depicted in Fig. \ref{fig_full_2}(a). As shown in the figure, the position of the main lobe shifts during propagation, and after propagation distance of $80\lambda$, an Airy-like intensity distribution emerges. The overall propagation trajectory is illustrated, where the beam initially focuses, then curves away from the center, and subsequently returns to the central region after a propagation distance of around $500\lambda$. Notice that each column is normalized to their own maxima, and the colomap only keeps value $>0.99$ to indicate the beam propagation trajectory in a simple form. The corresponding far-field radiation pattern is also given in Fig. \ref{fig_full_2}(a), the main lobe is around the broadside, with a slightly reduced gain compared to normal collimated Gaussian beam. Due to the lack of an accurate mathematical expression for the trajectory, optimizing the curved beam configuration to address specific blockage scenarios remains challenging. Nevertheless, recent progress with machine-learning-assisted strategies demonstrates promising potential for efficiently curving around obstacles \cite{Chen2024}.

\subsection{Amplitude and phase manipulation}\label{sec3.2}

\subsubsection{Anti-focus beam}\label{sec3.2.1}

The manipulation of both phase and amplitude also enables unique propagation behaviors. For example, in a scenario involving secure communications systems \cite{Jornet2023,Petrov20241}, two devices are positioned within the near-field region of the transmitter, as shown in Fig. \ref{fig_full_2}(b). The objective is for the closer device (the eavesdropper) to receive minimal power, while the farther device (the legitimate receiver) receives maximum power. This can be achieved using a so-called anti-focused beam, which, as its name implies, behaves oppositely to a focused beam. Instead of concentrating energy in a specific region, the anti-focused beam creates a zone with no power, entirely surrounded by regions of power in all three dimensions. This beam type is also known as a bottle beam \cite{Li2018} and is typically generated by the focusing property of a Bessel beam. Considering applying a focusing effect to a Bessel beam that consists of concentric rings, the boundary rings are focused to the first focal point, while the center rings are focused to the second focal point. Detailed principle can be explained by the ray tracing model shown in \cite{Wei2005}. \par

For example, considering a Bessel beam with a deflection angle $\theta = 13.8^\circ$, same as the one described in section \ref{sec3.1.6}, after a propagation of $45\lambda$, the Bessel beam is focused by a phase profile corresponding to a focal length of $20\lambda$. The resultant amplitude and phase distributions are shown in Fig. \ref{fig_full_2}(b). The beam propagation behavior is illustrated, where two distinct focal points are formed along the propagation axis, one in front of the null region and the other behind it, while maintaining a power-free zone in the intermediate region. Such unique propagation behavior can be confirmed by Fig. \ref{fig_full_2}(b). Such anti-focused beam characteristics offer potential advantages for secure communication by creating a power-free zone that reduces the risk of signal interception, thereby enhancing communications confidentiality.

\subsubsection{Laguerre-Gaussian beam}\label{sec3.2.2}

In the following scenario, we consider a case where the receiver requires maximum channel capacity, as shown in Fig. \ref{fig_full_2}(c). A promising method to enhance transmission speed without consuming additional spectral resources is mode multiplexing. These modes are orthogonal to each other, enabling simultaneous transmission of multiple independent data streams over the same frequency band. While previous sections primarily discussed near-field effects, it is important to note that mode orthogonality holds in both near- and far-field regions. These modes can be calculated by solving the Helmholtz equation under paraxial approximation. Under cylindrical coordinates, the complete solutions, incorporating both amplitude and phase, are described by Laguerre-Gaussian (LG) beams. These LG beams at $z = 0$ are mathematically expressed as \cite{Plick2015}:
\begin{equation}
LG_{l,p}(r, \varphi, 0) = C_{lp}^{LG} \left( \frac{\sqrt{2}r}{w_0} \right)^{|l|} \exp\left( -\frac{r^2}{w_0^2} \right) L_p^{|l|}\left( \frac{2r^2}{w_0^2} \right) \exp(jl\varphi),
\label{Eq:LG}
\end{equation}
where $l$ and $p$ represent the azimuthal and radial orders, $C_{lp}^{LG}$ is a constant term, and $ L_p^{|l|}$ is the  generalized Laguerre polynomials. A set of LG modes example is illustrated in Fig. \ref{fig_full_2}(c). \par 

A special case of the LG beams occurs when the radial order $p$ equals zero. In this case, mode orthogonality can be achieved through phase-only manipulation via their helical phase distribution, which is widely known for carrying orbital angular momentum (OAM) \cite{Su2023,Willner2025}. The helical phase distribution can be simplified as: $\phi = \mathrm{exp}(jl\varphi)$. An example of the phase distribution for $l = 2$ is shown in Fig. \ref{fig_full_2}(d). The corresponding near-field pattern at a propagation distance of $50\lambda$ and the cross sectional view are shown in Fig. \ref{fig_full_2}(d). Notably, A distinctive null at the center is observed, which is a characteristic feature caused by their phase singularity. Similarly, in the far-field pattern shown by Fig. \ref{fig_full_2}(d), the power null appears at the broadside, indicating that the receiver must either be positioned close to the transmitter or possess a sufficiently large aperture to capture the energy. Therefore, while mode orthogonality remains intact in the far-field, such a transmission system is less likely to function effectively in this region, presenting a significant practical implementation challenge. In the near-field region, a potential solution to mitigate the beam divergence effect is to apply a non-diffractive OAM beam \cite{Li2024_OAM}. Since OAM can be considered as an additional property introduced to other beams, the phase distribution can be viewed as the superposition of an OAM beam and a Bessel beam, allowing divergence to be suppressed within the non-diffractive region.

\subsubsection{Hermite-Gaussian beam}\label{sec3.2.3}

It is also possible to solve the Helmholtz equation under Cartesian coordinates, and the results can be expressed by Hermite-Gaussian (HG) beams. Hermite-Gaussian modes are a set of orthogonal solutions that describe the transverse field distribution of a Gaussian beam in terms of axial variations, and the Gaussian beam that was earlier discussed is the fundamental mode. HG modes are characterized by two indices: $m$ and $n$, which represent the number of nodes in the {\it x}- and {\it y}-directions, respectively. In Cartesian coordinate, these HG beams at $z = 0$ are mathematically expressed as \cite{Saghafi2001}:
\begin{equation}
HG_{m,n}(x, y, 0) = C_{mn}^{HG} {H_m}( \frac{\sqrt{2}x}{w_0}) {H_n}( \frac{\sqrt{2}y}{w_0}) \exp\left( -\frac{x^2 + y^2}{{w_0}^2} \right),
\label{Eq:HG}
\end{equation}
where $m$ and $n$ represent axial orders, $C_{mn}^{HG}$ is a constant term, and $H_m$ and $H_n$ are the Hermite polynomials. A set of HG modes example is illustrated in Fig. \ref{fig_full_2}(e). As shown, these modes exhibit symmetry between {\it x}- or {\it y}-axis, and due to the mutual orthogonality, these modes hold potential in high-speed communications systems as well. Similar to LG mode, power null can be observed and therefore practical implementation in the far-field region remain challenging.

\subsection{Comparison}\label{sec3.3}

\begin{table*}[!t]
\caption{{Comparison on different types of beam. NF: near-field. FF: far-field. NA: not applicable.}}
\begin{center}
\label{tab: Beam}
\begin{tabular*}{\textwidth}{@{\extracolsep\fill} p{2.1cm} p{3cm}<{\centering} p{2.8cm}<{\centering} p{2.1cm}<{\centering} p{1.8cm}<{\centering}}
\toprule 
Beam type & \makecell{Applications \\ in terahertz \\ communications}  &\makecell{NF and FF} & \makecell{Manipulation}  & {Complexity}  \\ 


\midrule 

\rowcolor{gray!30} 
{Focused} & \makecell{User requires \\ maximized energy} & \makecell{NF: focused \\ FF: reduced gain}  & \makecell{Phase} & {Low}  \\

\makecell[l]{Multi-\\focused} & \makecell{Multi-users for \\ maximized energy} & \makecell{NF: multi-focused \\ FF: NA } &  \makecell{Phase} & \makecell{Low} \\

\rowcolor{gray!30} 
\makecell[l]{Extended \\focal depth} & \makecell{Maximized energy \\ within a \\ potential range } & \makecell{NF: focused \\within a range \\ FF: reduced gain} & \makecell{Phase} & {Low} \\

Flat-top & \makecell{Maximized energy \\ coupling efficiency} & \makecell{NF: flat-top \\ FF: NA}   & \makecell{Phase} & {Low} \\

\rowcolor{gray!30} 
Arbitrary & \makecell{Holographic \\ communications} & \makecell{NF: desired \\ beam shape \\ FF: NA} & \makecell{Phase} & \makecell{Moderate} \\

Bessel & \makecell{Passing through \\ small obstacle} & \makecell{NF: non- \\diffraction region \\ FF: main lobe\\ at deflection \\ angle} &  \makecell{Phase} & {Low} \\

\rowcolor{gray!30} 
Airy & \makecell{Passing around \\ large obstacle} & \makecell{NF: curved \\ trajectory \\ FF: reduced gain. \\ Main lobe at \\ around broadside} &  \makecell{Phase} & {Moderate} \\

Anti-focused & \makecell{Confidential \\ communications} & \makecell{NF: power-free \\ zone surrounded \\ by power in all \\three dimensions \\ FF: NA } &  \makecell{Amplitude \\ and phase} & {High} \\

\rowcolor{gray!30} 
\makecell[l]{Laguerre- \\Gaussian} & {\makecell{Maximized channel \\ capacity}} & {\makecell{NF: mode \\ orthogonality \\ FF: NA due to \\power null}} &  \makecell{Amplitude \\ and phase} & \makecell{High} \\

\rowcolor{gray!30} 
\makecell[l]{(OAM)} &{$\sim$}& {$\sim$} &  \makecell{{(Phase)}} & \makecell{(Moderate)} \\


\makecell[l]{Hermite- \\Gaussian} & \makecell{Maximized channel \\ capacity} & \makecell{NF: mode \\ orthogonality \\ FF: NA due to \\power null} &  \makecell{Amplitude \\ and phase} & {High} \\

\bottomrule


\end{tabular*}
\end{center}
\end{table*}

A comparison of those aforementioned beam types is summarized in table \ref{tab: Beam}. Generally, phase-only manipulation results in lower complexity, except for some special cases such as arbitrary beam shapes or airy beams, which require rapid phase variation and may increase implementation difficulty. For OAM beams, achieving different modes for multiplexing might require a single phase plate, which can increase complexity as well. When both amplitude and phase manipulation are involved, the complexity rises significantly. The table also highlights potential applications in terahertz communications, and summarizes the effects in both near- and far-field conditions. \par

\section{Experiments of beam manipulation}\label{sec4}

In section \ref{sec3}, the concept of beam manipulation as well as the theoretical propagation behaviors are introduced. Building on that foundation, this section focuses on the experimental verification of beam manipulation through phase control, a widely used and efficient method in terahertz communications systems. For clarification, this section outlines the general principles and key steps involved in measurement, therefore presents only three typical cases rather than covering all the beam types discussed in the previous section. While various measurement systems are available and many research groups have conducted experiments on beam manipulation \cite{Guerboukha2024,Kludze2024,Abdellatif2024,Dechwechprasit2023}, the aim here is to provide general guidance for such processes rather than showing the state-of-art.  \par

In the microwave band, the phased array antenna is the most widely recognized method for beam manipulation. In this approach, power from the source is distributed to an antenna array via a feeding network, with each element incorporating a phase shifter to directly achieve the desired phase distribution. From an optics perspective, a spatial light modulator (SLM) provides a direct method for beam manipulation. SLMs can be adapted for terahertz frequencies by employing optical pumping to modulate the free carrier density in semiconductors, thereby altering their electromagnetic properties \cite{Li2022,Dechwechprasit2023a}. However, despite recent advancements, both phased array antennas and SLMs at terahertz frequencies still face practical limitations, primarily due to efficiency challenges \cite{Shrekenhamer2013,Abdo2022,Gao2024}. As a result, current practical free-space beam manipulation is predominantly implemented through static methods. A static approach utilizes passive devices to generate a fixed phase distribution tailored to a specific beam manipulation objective. This section focuses on the experimental setup for implementing passive lenses in such static beam manipulation scenarios. \par

\begin{figure}[!t]
\centering
\includegraphics[scale=0.69]{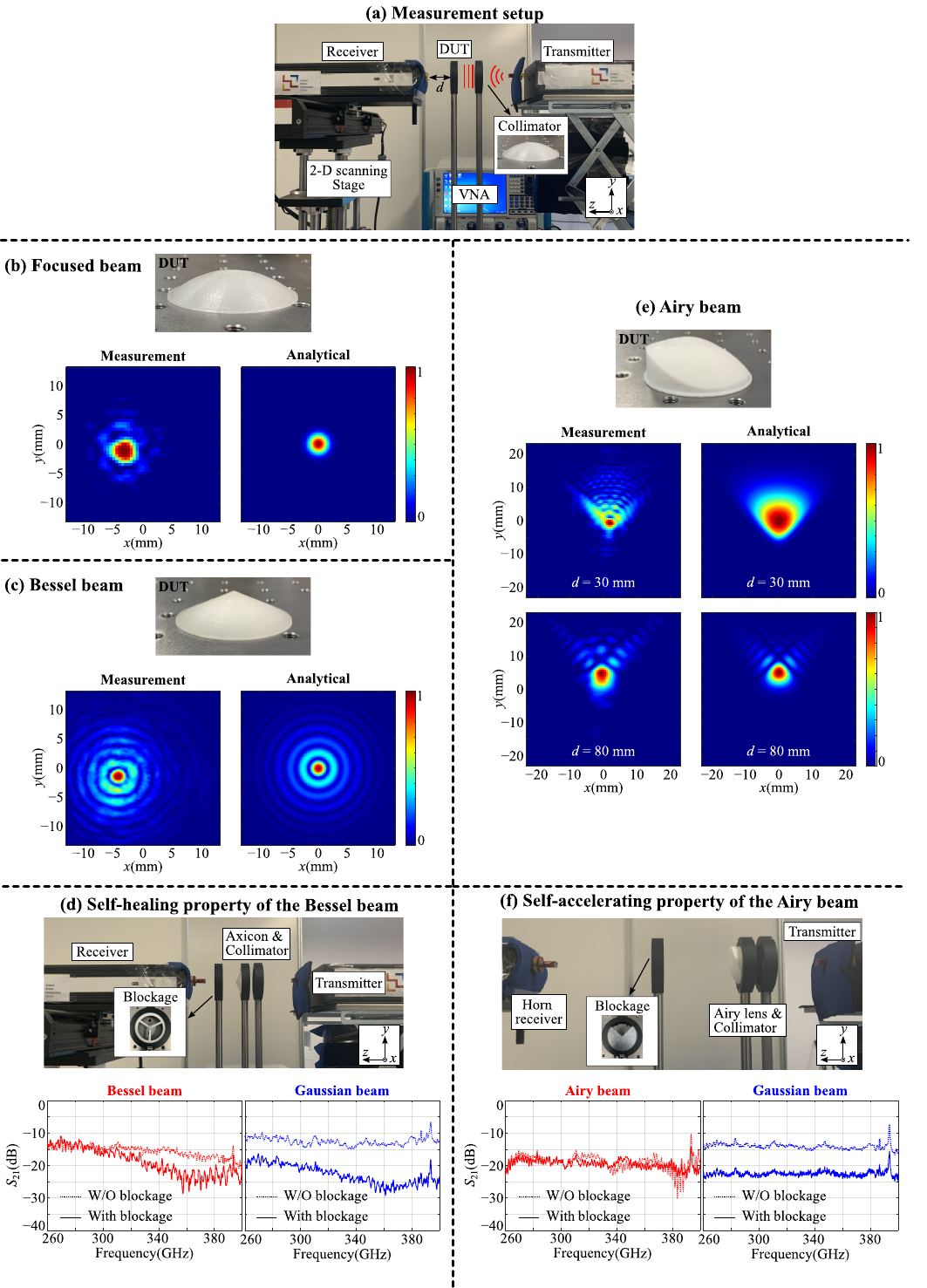}
\caption{Measurement setup and results. (a) A typical experimental setup for terahertz beam manipulation measurement. (b) Measurement results of a focused beam. (c, d) Measurement results of a Bessel beam and its self-healing property. (e) Measurement results of an Airy beam and its self-accelerating property.}
\label{fig_Exp_full}
\end{figure} 

Lens antennas operate based on traditional optics principles related to the path-length optics theory. As a wave propagates through a dielectric medium, it undergoes a phase delay that depends on the medium's refractive index $n_r$, the distance travelled {\it l}, and the wavenumber $k_0$. This phase delay is given by:
\begin{equation}
{\phi_\mathrm{delay}} = -k_0n_rl .
\label{Eq:Path_length}
\end{equation}
Consequently, phase manipulation can be achieved by modifying either the wave's travel path length or the refractive index of the medium. Conventional lens antennas control the phase by modifying the optical path length through their geometric design \cite{Llombart2011}. Planar lenses are also potential alternatives. These lenses maintain a fixed travel path length but achieve the desired phase variation by leveraging effective medium theory to create an artificial gradient-index (GRIN) \cite{Melendro-Jimenez2023,Li2024a}. \par

While material properties and fabrication methods play crucial roles in realizing practical terahertz beam manipulation \cite{Ako2020,Carter2023}, this section focuses on the implementation and experimental measurement of lens antennas. To simplify the demonstration, lens antennas are fabricated using 3-D printing technology. Although dielectric losses are inherent in 3-D printed materials, they are measured and compensated for in the experiments. Therefore, the impact of dielectric losses is neglected in the following discussion. \par

Figure \ref{fig_Exp_full}(a) illustrates a typical experimental setup for terahertz beam manipulation using a transmissive device at 300 GHz. The setup employs a Ceyear 3672C vector network analyzer (VNA), equipped with WR-2.8 extension modules covering the 260--400 GHz frequency range for precise terahertz signal measurement. The key modules at the transmitter and receiver sides are described as follows.\par

\begin{itemize}
\item{Transmitter side}: A WR-2.8 standard gain horn antenna is used to generate a quasi-Gaussian beam. After propagating approximately 50 mm, the beam forms a beam radius of around 10 mm at 300 GHz. At this point, the wavefront is no longer planar due to divergence. To ensure a uniform phase distribution, a collimator is typically introduced. \par

\item{Device under test (DUT)}: The fabricated lens is positioned adjacent to the collimator and is responsible for introducing the desired phase distribution to achieve the intended beam manipulation. Different DUTs will be discussed in subsequent sections.\par

\item{Receiver side}: The observation plane is situated at a distance $d$ from the DUT. The receiver consists of a WR-2.8 waveguide with a rectangular aperture of $0.711\times 0.356$ $\mathrm{mm}^2$. It should be pointed out that the waveguide's flange coupling may slightly affect its radiation pattern. For ideal performance, a near-field probe with nearly isotropic radiation pattern is recommended to ensure that the received signal is independent of the beam's incoming direction \cite{Li2023b,Chung2024}. The receiver is mounted on a two-axis motorized stage, enabling a maximum scanning area over $150\times 150$ $\mathrm{mm}^2$. The scanning resolution is set to match the diffraction limit and ensuring precise measurements.
\end{itemize}

\subsection{Focused beam}\label{sec4.1}

The phase profile $\phi_\mathrm{focus}$ obtained from section \ref{sec3.1.1} is unwrapped, and the lens structure is designed according to the formula:
$h = \frac{-\phi_\mathrm{focus}}{k_0 (n_r-n_0)}$, where $n_0$ is the refractive index of air equals to 1, and $n_r$ is the refractive index of the material. For commonly used 3-D printing materials such as polylactic acid (PLA) or cyclic olefin copolymer (COC), the refractive index for both material is approximately 1.5 \cite{Kristian2009,Brodie2022}. The lens is designed with a diameter of 50 mm, making it compatible with standard optical lens holders. The printed collimator sample is shown in Fig. \ref{fig_Exp_full}(b). In the experiment setup, one collimator is used to transform a spherical wavefront into a collimated beam, and an identical collimator as DUT focuses the beam at a focal length of approximately 50 mm.
The receiver is placed at 50 mm from the DUT, and measured beam profile at 300 GHz is shown in Fig. \ref{fig_Exp_full}(b), indicating the measured beam pattern closely matches the theoretical prediction. Minor sidelobes are observed, which are likely caused by measurement imperfections such as alignment errors.

\subsection{Bessel beam}\label{sec4.2}

Similarly, the phase profile described in section \ref{sec3.1.6} is unwrapped, and the lens structure designed with a deflection angle of $13.8^\circ$ is shown in Fig. \ref{fig_Exp_full}(c). In this configuration, one collimator is used to transform a spherical wavefront into a collimated beam, and an axicon as DUT to generate a Bessel beam. The measured field pattern at 300 GHz at an observation plane of around 30 mm is demonstrated in Fig. \ref{fig_Exp_full}(c), and the measured result agree well with the analytical calculations.

To further demonstrate the self-healing property of the Bessel beam, experiments are conducted with a small obstacle of approximately 5 mm in diameter, placed between the axicon and the receiver. The obstacle size is deliberately chosen to be smaller than the maximum non-diffraction width of the Bessel beam, which is around 10 mm, as explained in section \ref{sec3.1.6}. An aluminum foil is adhered to a printed structure to serve as the obstacle. Since this is a near-field operation, the horn antenna receiver is positioned close to the obstacle. Placing the receiver in the far-field would result in a significant reduction in received power due to a lower gain at broadside.

The experiment setup is shown in Fig. \ref{fig_Exp_full}(d). The measured $S_\mathrm{21}$ parameter is compared between the Bessel beam (with the axicon) and a normal collimated Gaussian beam (without the axicon). The obstacle results in a standing wave during propagation therefore strong fluctuations can be found in the measured $S_\mathrm{21}$. As shown in Fig. \ref{fig_Exp_full}(d), without any obstacle, the Bessel beam exhibits reduced received power of around 3 dB compared to the planar Gaussian beam. However, when the obstacle is introduced, the received power of the Bessel beam surpasses that of the collimated Gaussian beam. This outcome aligns with the expectation that, in terahertz communication scenarios, the Bessel beam's self-healing property can improve signal quality when small obstacles obstruct the transmission path.

\subsection{Airy beam}\label{sec4.3}

Following the required phase profile described in section \ref{sec3.1.7}, the 3-D printed lens designed to generate an Airy beam is shown in Fig. \ref{fig_Exp_full}(e). The measured field pattern at 300 GHz, observed at planes approximately 30 mm and 80 mm from the lens, is presented in Fig. \ref{fig_Exp_full}(e), respectively. The measured results align well with analytical calculations. As expected, as the propagation distance increases, the main lobe begins to shift upward, and adjacent sidelobes start to emerge.

To further demonstrate the self-accelerating property of the Airy beam, experiments were conducted with a large obstacle blocking around 75\% of the aperture, placed between the lens and the receiver. The blockage is positioned in a region where the main lobe of the airy beam is shifted towards the $+y$-direction, ensuring that the blockage has a negligible effect on the airy beam's propagation. Similarly, an aluminum foil is attached to a printed structure to partially block the transmission. The horn antenna receiver is positioned approximately 100 mm away from the obstacle. The measurement setup is given in Fig. \ref{fig_Exp_full}(f). The measured transmission coefficients are compared between the Airy beam (with the airy lens) and the collimated Gaussian beam (without the airy lens) are shown in Fig. \ref{fig_Exp_full}(f). For the Gaussian beam, introducing such an obstacle leads to a power reduction of over 6 dB. However, for the Airy beam, although the received power is slightly reduced, the introduction of the obstacle results in almost no change to the received power. This supports the expectation that, in terahertz communication scenarios, the Airy beam's curved trajectory allows it to bypass large obstacles that would otherwise block the transmission path.

\section{Metasurface towards reconfigurability}\label{sec5}

Although passive lenses are the most cost-effective solutions for beam manipulation in the terahertz range, their bulkiness and intrinsic dielectric losses significantly hinder their practical implementation in terahertz communication systems. In contrast, metasurface technology has emerged as a promising alternative for beam manipulation due to its low-profile design and high integrability. The measurement principles for metasurfaces are generally similar to those for lens antennas, with the setup replacing the lens antenna with the proposed transmissive metasurface. For reflective metasurfaces, the setup differs slightly, as can be explored in references \cite{Qu2017,You2019}.

\subsection{Metasurface}\label{sec5.1}

Metamaterials, made up from sub-wavelength unit cells, with their ability to effectively manipulate electromagnetic waves in terms of amplitude, phase, and polarization \cite{Medrar2021,Quader2024,You2023}. Each unit cell interacts with EM waves behaves like electrical components, such as capacitors, inductors, and resistors. For example, capacitive coupling arises from the separation of conductive patches, inductance is associated with current flow along conductive paths, and resistance accounts for energy loss due to absorption or intrinsic material properties. These components collectively form an equivalent circuit model, explaining the metasurface's transmission or reflection characteristics, which can be utilized for phase manipulation. \par

Metasurfaces can be fabricated using a variety of materials to achieve diverse functionalities. For instance, dielectric silicon resonators can serve as reflect arrays, enabling quarter-wave or half-wave mirrors \cite{Lee2018}, or as absorbers in specific frequency ranges \cite{You2020}. Hybrid designs, such as thin patterned gold films deposited on dielectric substrates, expand the design possibilities for transmit arrays with tailored phase profiles \cite{Chang2017,You2021,You2020a,Shiri2023}. In metasurface lens designs, precise phase control at the unit-cell level is critical, requiring a phase response that spans at least $2\pi$ to construct a discrete Fresnel-equivalent lens. These features make metasurfaces highly promising for terahertz applications, offering significant design flexibility and performance enhancements. For more reference related to the concept, design, fabrication and applications of terahertz metasurface, we direct interested readers to the following references \cite{Al-Naib2017,Ako2020,You2022,You2024}.

\subsection{Reconfigurable intelligent surface}\label{sec5.2}

Still building on the metasurface principle but one step forward, active semiconductor components or tunable materials can be integrated into the unit cell structure, leading to the design of reconfigurable intelligent surfaces (RIS), also known as programmable metasurfaces. Since these active components require power inputs and wiring connections, they are typically implemented in a reflective mode rather than transmissive. In this article we briefly review two common types of RIS, namely, semiconductor-based and tunable material-based. For more detailed discussions on these types of RIS, or other types based on micro-electromechanical system (MEMS), mechanical tuning and space-time modulation, we recommend consulting the referenced materials \cite{Fu2020,Rasilainen2023,Fu2024,Wu2022,Wu2024}.  \par 

On one hand, semiconductor-based metasurfaces utilize active components such as diodes and transistors. In principle, applying a bias voltage to diodes, such as PIN and varactor diodes, adjusts the resistance or capacitance of the metasurface elements, therefore achieves a different phase response and enabling real-time phase modulation \cite{Su2015,Liu2023a}. Transistors, such as high-electron-mobility transistors (HEMTs), which operate more efficiently at terahertz frequencies, enable a better stability and a high-speed modulation, making them suitable for terahertz communication systems \cite{Zhang2019,Pan2021,Lan2023}. Additionally, semiconductor-based metasurfaces can be developed into on-chip configurations, offering more compact, highly integrated solutions for ultrafast modulation in terahertz systems. These on-chip metasurfaces integrated with tunable components are also suitable for applications in sensing and imaging applications. Notable advancements include a 300 GHz tiled CMOS chip enabling dynamic beamforming across a $\pm 30^{\circ}$ range \cite{Venkatesh2020} and a $98\times 98$ unit CMOS array with a $1^\circ$ beamwidth at 265 GHz \cite{Monro2022}. \par

On the other hand, tunable material-based metasurfaces leverage advanced materials like liquid crystals (LC) and vanadium dioxide ($\mathrm{VO}_2$) to modulate electromagnetic properties. Liquid crystals, with their birefringence properties under varying bias voltages, i.e., the equivalent dielectric constant of the LC, and therefore the phase response, can be changed according to the voltage applied. Such tunable behavior enables precise phase control in applications such as high-resolution beam scanning, with a low cost and well-established fabrication processes \cite{Fu2023,RIS_Wang2024}. $\mathrm{VO}_2$, due to its unique material characteristics, is capable of switching between insulating and metallic states through electrical,optical, or thermal stimulation. The insulator-metal transition property also provides versatile and cost-effective solution for dynamically tunable phase distribution \cite{Hashemi2016,Chen2022}.

\subsection{Comparison}\label{sec5.3}

While metasurface, RIS technologies, as well as the lens designs discussed in section \ref{sec4} have all shown effective beam manipulation capabilities, it is important to recognize that each method has its own advantages and limitations. The selection of an appropriate approach depends on specific conditions and requirements. To facilitate this, we provide a general comparison of these three technologies in table \ref{tab: Manipulation}.  \par

Lens designs generally offer a cost-effective solution for static beam manipulation, as they can be fabricated using commercially-available 3-D printing technology or CNC machining. These printed lenses provide continuous phase profiles, ensuring accurate beam manipulation performance in most cases. Given that most materials can be considered as dispersionless within the communications bandwidths, the resulting phase profile is linearly proportional to the wavenumber and the lens designs are conceptually achromatic. A key limitation of lens designs is their large device size, which can lead to increased intrinsic material losses. Consequently, the aperture size of the lens design cannot be too large, and the overall compactness is compromised. Introducing phase wrapping can significantly reduce the device profile, but there is a trade-off with a decrease in bandwidth. \par

To improve compactness, metasurfaces provide an alternative solution. They involve advanced fabrication techniques such as thin-film spin coating or electron beam deposition, making the fabrication process more complicated than lens devices. The unit cells of a metasurface are required to cover at least a $2\pi$ phase range, and a 3-bit configuration is typically sufficient. For example, eight unit cells are designed to cover the $2\pi$ phase range, with each unit cell contributing a phase shift of approximately $45^\circ$. While more precise phase manipulation using higher-order bit configurations is possible, it introduces additional complexity. The required phase profile is discretized, and the unit cells are mapped to the phase profile accordingly. This results in relatively good beam manipulation performance, although discretization might slightly limit the precision. Due to the necessary phase wrapping and intrinsic dispersion where the phase difference between unit cells does not remain constant across a broad frequency band, the bandwidth performance is not as large as that of lens designs. However, it is typically sufficient for most communications applications. Metasurfaces are designed with tailored impedance matching, therefore minimizing radiation losses and ensuring good efficiency. A challenge with metasurfaces is that the amplitude response is also coupled to the phase response. For example, the amplitude responses of the eight unit cells covering the $2\pi$ phase range might be slightly different, which can lead to complexity in designing a particular wavefront. Nonetheless, there is no strict limitation on the aperture size, allowing for the design of very large metasurfaces. In summary, metasurfaces offer a more compact solution than lenses with acceptable performance, but they are still designed for one specific goals and do not offer reconfigurability. \par

\begin{table*}[!t]
\caption{\justifying{Comparison on different beam manipulation strategies. No specific numerical values are presented here, since the types and characteristics of the strategies can vary significantly depending on the context and conditions. This comparison aims to highlight the general trends rather than provide precise data.}}
\begin{center}
\label{tab: Manipulation}

\begin{tabular*}{\textwidth}{@{\extracolsep\fill} p{2.7cm} p{3cm} p{3cm} p{3cm}}


\toprule 
Type & \makecell{Lens}  & \makecell{Metasurface} & \makecell{RIS}  \\ 

\midrule 

\rowcolor{gray!30} \makecell[l]{Phase \\ quantization} & \makecell{{Continuous}} & \makecell{Discretized \\ (at least 3-bit)}  & \makecell{Discretized \\ (typically 1-bit)}  \\

Bandwidth & \makecell{Large \\ (Achromatic level) } & \makecell{Moderate \\ (Limited due to \\ dispersion)} & \makecell{Small \\ (Oscillation level)}  \\

\rowcolor{gray!30} 
Efficiency & \makecell{Moderate \\ (Loss depends on \\ bulkiness)} & \makecell{{High} \\ (Tailored matching \\ impedance)} & \makecell{Low \\ (Loss due to \\  less efficient active \\ components)} \\

\makecell[l]{Aperture size} & \makecell{Moderate \\ (Limited by \\bulkiness)} & \makecell{{Large} \\ (No strict \\limitation)} & \makecell{Small\\ (Limited by \\ manufacturing)} \\

\rowcolor{gray!30} 
Compactness & \makecell{Low} & \makecell{Moderate} & \makecell{{High}} \\

Reconfigurability & \makecell{Static} & \makecell{Static} & \makecell{{Dynamic}} \\

\rowcolor{gray!30} 
Cost & \makecell{{Low}} & \makecell{Moderate} & \makecell{High} \\

\bottomrule


\end{tabular*}
\end{center}
\end{table*}

Reconfigurable intelligent surfaces (RISs) are promising solutions for enabling reconfigurability, as the phase response of their unit cells can be dynamically adjusted, allowing for the realization of various beam manipulation goals. However, enabling such reconfigurability also introduces several constraints. Take semiconductor-based RIS as an example, the phase response is controlled by switching the active components on and off. As a result, the phase profile is typically discretized into a 1-bit stage, meaning the phase can only take values of $0^\circ$ or $180^\circ$. Consequently, the performance of RIS is generally not as good as lens or metasurface strategies. Furthermore, the phase difference between the on and off states of the active components can maintain at $180^\circ$ for a narrow frequency range, limiting the flexibility of RIS to operate at broader frequencies. In the case of tunable material-based RIS, multi-bits phase quantization is also possible, with an improved bandwidth performance. However, these RISs face challenges such as slow switching speeds, often on the millisecond scale, which limits their practical applications. The efficiency of different RISs is typically lower than that of lens or metasurface designs, due to introduction of active components or tunable materials that generates significant losses. A critical limitation of RISs is that the current state of the art does not support very large aperture sizes. The fabrication complexity of RISs is extremely high, requiring power inputs and wiring connections, which restricts the aperture size compared to static approaches. As a result, RISs are less suitable for large-scale applications. In summary, RISs offer a compact solution for achieving reconfigurability in beam manipulation, but at the cost of reduced performance and higher complexity. Despite these limitations, RISs play a crucial role in the future development of terahertz communication systems, especially as technologies for dynamic beam manipulation continue to evolve.

\section{Conclusion}\label{sec6}

This article provides an overview of techniques for achieving highly-efficient beam manipulation in the terahertz range. We laid out the fundamental principles of electromagnetic wave propagation, followed by detailed beam manipulation techniques that address various practical challenges in terahertz communication scenarios. The core principle of beam engineering, manipulating the complex electric field by adjusting the aperture’s amplitude and phase distribution to achieve desired beam behavior, is discussed in detail. To showcase a simple and cost-effective approach, experimental validations using 3-D printed lens solutions are presented, demonstrating results consistent with theoretical predictions. Alternative methods, such as metasurfaces and reconfigurable intelligent surfaces, are also briefly explored. \par

The primary contribution of this article is to provide comprehensive guidance on beam manipulation for the terahertz communications community, detailing critical steps such as principles, design, fabrication, and measurement processes to realize cross far- and near-field coverage. It is important to acknowledge that the beam manipulation techniques presented here remain constrained by the current state of fabrication and implementation technologies. While a variety of scenarios are explored, they are not exhaustive. Looking ahead, the advancement of solid-state terahertz sources with dynamic beam manipulation capabilities, potentially implemented through efficient phased-array antennas, will be critical for future terahertz systems. These systems will require increasingly complicated beam manipulation techniques. The evolution of these approaches must align with the progress in fabrication technologies. Successful implementation will profoundly enhance productivity, paving the way for high-quality productive forces.






\bibliography{sn-article}

\end{document}